\documentstyle[epsfig,aps]{revtex}
\begin{document}
\draft
\title{Universal canonical entropy for gravitating 
systems\footnote{Based on talk delivered at the International 
Conference on Gravitation and Cosmology 2004, held at Kochi, India 
during 5-10 January, 2004.}} 
\author{Ashok Chatterjee\footnote{email: 
ashok@theory.saha .ernet.in} and Parthasarathi Majumdar\footnote{On
deputation from the Institute of Mathematical Sciences, Chennai 600
113, India; email: partha@theory.saha.ernet.in}}
\address{Theory Group, Saha Institute of Nuclear Physics, Kolkata 700 
064, India.}   
\maketitle
\begin{abstract}
The thermodynamics of general relativistic systems with boundary,
obeying a Hamiltonian constraint in the bulk, is argued to be
determined solely by the boundary quantum dynamics, and hence by
the area spectrum. Assuming, for large area of the boundary, (a) an
area spectrum as determined by Non-perturbative Canonical Quantum
General Relativity (NCQGR), (b) an energy spectrum that bears a
power law relation to the area spectrum, (c) an area law for the
leading order microcanonicai entropy, leading thermal fluctuation
corrections to the canonical entropy are shown to be logarithmic in
area with a universal coefficient. Since the microcanonical entropy
also has univeral logarithmic corrections to the area law (from
quantum spacetime fluctuations, as found earlier) the canonical
entropy then has a universal form including logarithmic corrections
to the area law. This form is shown to be independent of the index
appearing in assumption (b). The index, however, is crucial in
ascertaining the domain of validity of our approach based on
thermal equilibrium.  
\end{abstract}

\section{Introduction}

The asymptotically flat Schwarzschild spacetime is well-known
\cite{hawp} to have a thermal instability: the Hawking temperature
for a Schwarzschild black hole of mass $M$ is given by $T \sim 1/M$
which implies that the specific heat $C \equiv \partial 
M/\partial T <0 ~!$ The instability is attributed, within a 
standard canonical 
ensemble approach, to the superexponential growth of the 
density of states  $\rho(M) \sim \exp M^2 $ which results in the 
canonical partition function diverging for large $M$. 

The problems with an approach based on an equilibrium canonical 
ensemble do not exist, at least for isolated spherically 
symmetric black holes,
formulated as {\it isolated horizons} \cite{aa1} of fixed horizon 
area; these can be
consistently described in terms of an equilibrium {\it
microcanonical} ensemble with fixed ${\cal
A}$ (and hence disallowing thermal fluctuations of the energy $M$). 
For ${\cal A} >> l_{Planck}^2$, it has
been shown using Loop Quantum Gravity \cite{aa2}, that all
spherically symmetric four dimensional isolated horizons possess a
microcanonical entropy obeying the Bekenstein-Hawking Area Law
(BHAL) \cite{bek},\cite{haw}. Further, the microcanonical entropy
has corrections to the BHAL due to quantum spacetime fluctuations
at fixed horizon area. These arise, in the limit of large ${\cal
A}$, as an infinite series in inverse powers of horizon area 
beginning with a term logarithmic in the area \cite{km}, with 
completely finite coefficients,
\begin{eqnarray}
S_{MC} =S_{BH} -\frac32 \log S_{BH} + const.+ {\cal
O}(S_{BH}^{-1}) ~.\label{smc}
\end{eqnarray}
where $S_{BH} \equiv {\cal A}/4 l_{Planck}^2$

On the other hand, asymptotically anti-de Sitter (adS) black holes
with spherical symmetry are known \cite{hawp} to be describable in
terms of an equilibrium canonical ensemble, so long as the
cosmological constant is large in magnitude. For this range of
black hole parameters, to leading order in ${\cal A}$ the canonical
entropy obeys the BHAL. As the magnitude of the cosmological
constant is reduced, one approaches the so-called Hawking-Page
phase transition to a `phase' which exhibits the same thermal
instability as mentioned above.

In this talk, we focus on the following
\begin{itemize}
\item Is an understanding of the foregoing features of 
black hole entropy on some sort of a `unified' basis possible ? We 
shall argue, following \cite{cm} that it is indeed so, with some 
rather general assumptions.
\item In addition to corrections (to the area law) due to fixed 
area quantum spacetime fluctuations computed using a microcanonical 
approach, can one compute corrections due to {\it 
thermal} fluctuations of horizon area within the canonical ensemble 
? Once again, the answer is in the affirmative with some caveats. The 
result found in \cite{cm}, at least for the leading $\log area$ 
corrections, turns out 
to be {\it universal} in the sense that, just like the BHAL, it 
holds for all black holes independent of their parameters. 
\end{itemize}

\section{Canonical partition function : holography ?}

Following \cite{pm}, we start with the canonical partition function 
in the quantum case
\begin{eqnarray}
Z_C(\beta)~=~Tr~\exp -\beta \hat{H} ~.
\label{qpf}
\end{eqnarray}
Recall that in classical general relativity in the Hamiltonian
formulation, the bulk Hamiltonian is a first class constraint, so 
that
the entire Hamiltonian consists of the boundary contribution $H_S$
on the constraint surface. In the quantum domain, the Hamiltonian
operator can be written as
\begin{eqnarray}
\hat{H}~=~\hat{H}_V~+~\hat{H}_S~,
\label{deco}
\end{eqnarray}
with the subscripts $V$ and $S$ signifying bulk and boundary terms
respectively. The Hamiltonian constraint is then implemented by 
requiring
\begin{eqnarray}
\hat{H}_V~|\psi \rangle_V~=~0~
\label{hcon}
\end{eqnarray}
for every physical state $|\psi \rangle_V$ in the bulk. Choose as 
basis for the Hamiltonian in (\ref{deco}) the state $|\psi 
\rangle_{blk} \otimes |\chi\rangle_{bdy}$. This implies that the 
partition function may be factorized as
\begin{eqnarray}
Z_{C}~&\equiv&~ Tr \exp -\beta {\hat H} \nonumber \\
&=&~\underbrace{dim~{\cal 
H}_{bulk}}_{{indep.~of~\beta.}}~\underbrace{Tr_{bdy}
\exp -\beta {\hat H}_{bdy}}_{{boundary}} \label{facto}
\end{eqnarray}
Thus, the relevance of the bulk physics seems rather limited due to  
the constraint (\ref{hcon}). The partition function further reduces 
to
\begin{eqnarray}
Z_C(\beta)~=~dim~{\cal H}_V~Z_S(\beta)~,
\label{redp}
\end{eqnarray}
where ${\cal H}_V$ is the space of bulk states $|\psi\rangle$ and 
$Z_S$ is the `boundary' partition function given by
\begin{eqnarray}
Z_S(\beta)~=~Tr_S~\exp -\beta \hat{H}_S~.
\label{zs}
\end{eqnarray}
Since we are considering situations where, in addition to the
boundary at asymptopia, there is also an inner boundary at the
black hole horizon, quantum fluctuations of this
boundary lead to black hole thermodynamics. The factorization in
eq.(\ref{redp}) manifests in the canonical entropy as the
appearance of an additive constant proportional to $dim~{\cal
H}_V$. Since thermodynamic entropy is defined only upto an additive
constant, we may argue that the bulk states do not play
any role in black hole thermodynamics. This may be thought of as the 
origin of a weaker version of the holographic hypothesis \cite{thf}.

For our purpose, it is more convenient to rewrite (\ref{zs}) as
\begin{eqnarray}
Z_C(\beta)~~=~~\sum_{n \in {\cal Z}}
\underbrace{g\left (E_{bdy}({\cal A}(n)) \right)}_{{degeneracy}}~\exp 
-\beta E_{bdy}({\cal A}(n))~, \label{zss}
\end{eqnarray}
where, we have made the assumptions that (a) the energy is a function 
of the area of the horizon ${\cal A}$ and (b) this area is quantized. 
The first assumption (a) basically originates in the idea in the last 
paragraph of that black hole thermodynamics ensues solely from the 
boundary states whose energy ought to be a function of some property 
of the boundary like area. The second assumption (b) is actually 
explicitly provable in theories like NCQGR as we now briefly digress 
to explain. 

\section{Spin network basis in NCQGR}

The basic canonical degrees of freedom in NCQGR are holonomies of a
distributional $SU(2)$ connection and fluxes of the densitized triad
conjugate to this connection. The Gauss law (local $SU(2)$
invariance) and momentum (spatial diffeomorphism) constraints are
realized as self-adjoint operators constructed out of these
variables. States annihilated by these constraint operators span the
kinematical Hilbert space. Particularly convenient bases for this
kinematical Hilbert space are the spin network bases. In any of these
bases, a (`spinet') state is described in terms of {\em links} $l_1,
\dots, l_n$ carrying spins ($SU(2)$ irreducible representations)
$j_1, \dots j_n$ and {\it vertices} carrying invariant $SU(2)$ 
tensors 
(`intertwiners'). Spacetime curvature has support only on network. a 
particularly important property of such bases is that geometrical 
observables like area operator is diagonal in this basis with 
{\em discrete spectrum}. An internal boundary of a spacetime
like a horizon appears in this kinematical description as a  
punctured ${\cal S}^2$ with each
puncture having a deficit angle $\theta=\theta(j_i), i=1,\dots,p$, as 
shown in Fig.1. 
\begin{center}
\epsfxsize=60mm
\epsfbox{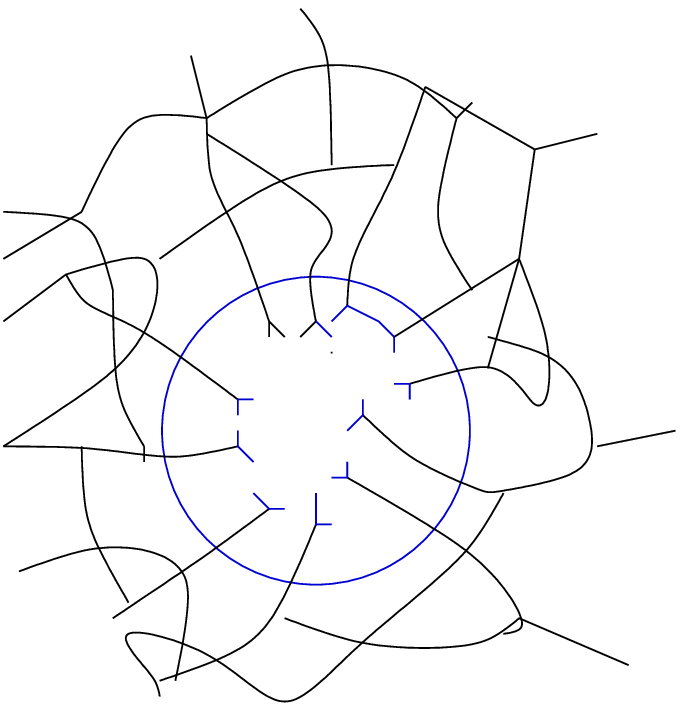}

{Fig. 1 {\em Internal boundary (horizon) pierced by spinet 
links}}
\end{center}
For macroscopically large boundary areas ${\cal A} >> l_{Planck}^2$, 
the area spectrum is dominated by $j_i=1/2, \forall i=1, \dots,p,~ p 
>>1$. This is the situation when the deficit angles at each puncture 
takes its smallest nontrivial value, so that a classical horizon 
emerges. That implies that 
\begin{eqnarray}
{\cal A}_{bdy}(p) ~\sim ~p~ l_{Planck}^2~,~p~>>~1  ~. \label{area}
\end{eqnarray}
This completes our digression on NCQGR. 

\section{Fluctuation effects on canonical entropy}

Going back to eq. (\ref{zss}), we can now rewrite the partition 
function as an integral, using the Poisson resummation formula
\begin{eqnarray}
\sum_{n=-\infty}^{\infty} f(n)~=~\sum_{m=-\infty}^{\infty}
\int_{-\infty}^{\infty} dx ~\exp (-2\pi i mx)~f(x)~. \label{poi}
\end{eqnarray}  
For macroscopically large horizon areas ${\cal A}(p)$, $x >> 1$, 
so that the summation  on the {\em rhs} of (\ref{poi}) is dominated 
by 
the contribution of the $m=0$ term. In this approximation, we have
\begin{eqnarray}
Z_C &~\simeq~& \int_{-\infty}^{\infty} dx~g(E(A(x)))~\exp -\beta
E(A(x)) \nonumber \\
&=&~\int dE~\exp [S_{MC}(E)~-~\log|{dE \over dx}|~-~\beta E]
\label{appr}
\end{eqnarray}
where $S_{MC} \equiv \log g(E)$. 

Now, in equilibrium statistical mechanics, there is an inherent
ambiguity in the definition of the microcanonical entropy, since it
may also be defined as ${\tilde S}_{MC} \equiv \log \rho(E)$ where
$\rho(E)$ is the density of states. The relation between these two 
definitions involves the `Jacobian' factor $|dE /dx|^{-1}$
\begin{eqnarray} 
{\tilde S}_{MC}~=~S_{MC}~-~\log|{dE \over dx}| ~.\label{amb}
\end{eqnarray}
Clearly, this ambiguity is irrelevant if all one is interested in is 
the leading order BHAL. However, if one is interested in logarithmic 
corrections to BHAL as we are, this difference is crucial and must be 
taken into account. 

We next proceed to evaluate the partition function in eq. 
(\ref{appr}) using the saddle point approximation around the point 
$E=M$ where $M$ is to be identified with the (classical) mass of the 
boundary (horizon). Integrating over the Gaussian fluctuations around 
the saddle point, and dropping higher order terms, we get,
\begin{eqnarray}
Z_C &\simeq& \exp {\{ S_{MC}(M) - \beta M - \log|{dE \over
dx}|_{E=M}] \}} \nonumber \\
& \cdot & \left [{\pi \over -S_{MC}''(M)} \right]^{1/2}~.
\label{spa}
\end{eqnarray}
Using $S_C = \log Z_C + \beta M$, we obtain for the canonical 
entropy $S_C$
\begin{eqnarray}
S_C~&=&~S_{MC}(M)~\underbrace{-~\frac12
\log(-\Delta)}_{{\delta_{th}S_C}}~, \label{scan}
\end{eqnarray}
where, 
\begin{eqnarray}
\Delta~\equiv~{d^2 S_{MC} \over dE^2}~\left({dE \over
dx}\right)^2|_{E=M}~. \label{delta}
\end{eqnarray}
Eq. (\ref{scan}) exhibits the equivalence of the microcanonical and 
canonical entropies, exactly as one expects when thermal fluctuation 
corrections are ignored. A few elementary manipulations on 
(\ref{delta}) yield
\begin{eqnarray}
\Delta~=~\left[ {d^2 S_{MC} \over d{\cal A}^2}~-~\left({dS_{MC} \over
d{\cal A}}\right) \underbrace{{d^2 E/d{\cal A}^2 \over
dE/d{\cal A}}}_{{non-univ.}}
\right]~\left({d{\cal A} \over dx}\right)^2|_{E=M}~. \label{delt}
\end{eqnarray}
Here, we observe that the microcanonical entropy obeys the BHAL {\it 
universally}, i.e., independent of the horizon parameters, and may 
even have universal logarithmic corrections in the horizon area. 
However, the factors in the {\it rhs} of (\ref{delt}) underbraced 
`non-univ.' depend explicitly on the area dependence of the energy 
and is hence a function of the horizon parameters. 

We now make the following assumptions
\begin{itemize} 
\item Assume $E({\cal A})~=~const.~{\cal A}^n$
\item Assume $S_{MC}({\cal A})~\sim~{\cal A}$
\end{itemize}
Recall also that ${\cal A}~\sim~x~for~x >> 1 
(large~area)$. Substitution in eq. (\ref{delt}) now leads to the 
following simple formula
\begin{eqnarray}
\delta_{th}S_C~=~ \frac12 \log S_{BH} ~-~ \frac12
\log(n-1)~+~const~.\label{ther}
\end{eqnarray}
The thermal fluctuation correction to the canonical
entropy of a spacetime with an inner boundary is {\bf universal},
independent of $r$; it is also insensitive (for large areas) to the
log(area) corrections in the microcanonical entropy due to quantum
spacetime fluctuations. We note en passant that the contribution 
due to $dE/dx$ was first 
included in the canonical ensemble in ref. \cite{gm}, although not on 
the grounds we have included it in this paper. An earlier paper by us 
\cite{cm2}
which delineated the contribution of fixed-area quantum spacetime 
fluctuations to the BHAL as distinct from thermal fluctuation effects 
missed out this `Jacobian' term. Similar is the case with 
ref.\cite{dmb} which also used the saddle point approximation to 
express the microcanonical entropy in terms of the canonical 
entropy.\footnote{Expressing $S_{MC}$ in terms of $S_C$, as in 
\cite{dmb}, may be 
technically advantageous on occasion, but conceptually difficult to 
comprehend, since the former excludes all energy fluctuations which 
the latter incorporates. Thus, when $S_{MC}$ is expressed in 
terms of $S_C$, the $\frac12 log {\cal A}$ correction 
we have found merely cancels an equal and opposite term in $S_C$ in 
the treatment of \cite{dmb}. Furthermore, it is hard to conclude from 
this treatment the fact that asymptotically flat black hole 
spacetimes, with fixed horizon area, can well be in 
equilibrium within a {\it microcanonical} ensemble.}

Rcalling that there is at least `circumstantial' evidence that the
microcanonical entropy has a `universal' form \cite{car}-\cite{sen},
identical to that obtained in ref. \cite{km} quoted in eq.
(\ref{smc}), the total canonical entropy, including both the
`finite-size' logarithmic corrections due to quantum spacetime
fluctuations and the thermal energy (area) fluctuations, is given by 
\begin{eqnarray}
S_C ~=~ S_{BH} ~-~ \log S_{BH} ~-~ \frac12 \log(n-1) ~+~ const. 
~+~\dots 
~.\label{sct}
\end{eqnarray}
Clearly, the nature of the quantum and thermal fluctuations 
corrections preseves the desired property of superadditivity for the 
canonical entropy. 

Two remarks are in order at this point: first of all, the form of the
$S_{MC}$ quoted in eq. (\ref{smc}) arises from the assumption that the
residual gauge subgroup of local Lorentz ($SL(2,C)$ invariance,
which survives on the Cauchy slice of the isolated horizon, is assumed
to be $SU(2)$. The log correction found in \cite{km} originates from 
counting the boundary states of an $SU(2)$ Chern Simons theory, where 
the boundary has the topology of a punctured $S^2$. On the other hand, 
it has been argued \cite{aa1} that the boundary conditions appropriate 
to a
non-rotating isolated horizon leave only a compact $U(1)$ subgroup 
of this $SU(2)$ on the Cauchy slice of the isolated horizon. If so, 
the $-(3/2) \log S_{BH}$ in (\ref{smc}) is replaced by $-(1/2) \log 
S_{BH}$ \cite{km2}, \cite{pm}, \cite{gmi}, and consequently there is 
{\it no} log(area) correction to 
the canonical entropy, as the thermal fluctuation correction
precisely cancels the quantum spacetime fluctuation correction. This 
is reminiscent of the phenomenon of `non-renormalization' that often 
occurs in certain quantum field theories. This cancellation is 
universal in the sense that it should hold for all non-rotating 
isolated horizons and presumably for rotating ones as well. 

The second remark pertains to the role of the index $n$ appearing in
the assumption regarding the power law relation between the boundary
energy and boundary area. Observe that the area dependence of the 
correction term is quite independent of $n$. However. $n$ has a 
crucial role to play: it determines the range of validity of the 
saddle point approximation used to evaluate $Z_C$. For $n>1$, both 
$S_C$ and the canonical Gibbs free energy remain real, implying that 
the saddle point has been correctly found. For $n<1$, however, both 
the canonical entropy and the free energy acquire an imaginary piece, 
signifying a {\it breakdown} of the saddle point approximation. 
Forcing the saddle point at the value $M$ implies that this is a point 
of {\it unstable} equilibrium. Thus, for example, using the formulas 
worked out in \cite{cm2}, \cite{dmb}, we obtain
\begin{itemize}
\item $n=2$ for the non-rotating 2+1 dimensional BTZ black hole so 
long as the horizon radius $r_H > (-\Lambda)^{-1/2}$, 
signifying the validity of the calculation in this case;
\item $n=3/2$ for the four dimensional adS Schwarzschild black hole, 
also indicating that the formula is reliable for this case in the 
same range of $r_H$. But,
\item $n=1/2$ for the four dimensional asymptotically flat 
Schwarzschild black hole, delineating the thermal instability 
mentioned at the very outset. 
\end{itemize}
The thermal instability for $n<1$ is generic for {\it all}
asymptotically flat (and also dS) black holes at large area, including 
the extremal limits. On the other hand, for the adS black holes, so 
long as one stays away from the Hawking-Page phase transition to the 
adS gas phase, there is no thermal instability \cite{hawp}. 

\section{Conclusions}

The canonical entropy of gravitating systems with horizons has a
universal correction to the area law due to thermal fluctuations, in
the form of $1/2~\log {\cal A}$, provided certain very general
assumptions are made about the relation between the energy and the
area of the boundary. Within these assumptions, asymptotically flat
black holes display an unstable thermal equilibrium exactly as
expected on general grounds. Inclusion of the finite-size quantum
corrections, in a microcanonical ensemble corresponding to fixed
boundary area, leads either to a net logarithmic correction to the
area law for canonical entropy obeying superadditivity, if the gauge
group on the boundary is $SU(2)$, or to {\it no} net logarithmic
correction at all, if the gauge group is compact $U(1)$.

At the top of the agenda for future work, then, is the investigation
of the assumptions made, within the framework a theory of quantum
gravitation like NCQGR.  One expects to have a better understanding of
the thermal instability encountered for asymptotically flat (and also
dS) black holes form such an exploration.

\end{document}